\documentclass[prd,superscriptaddress,showpacs,nofootinbib,amsmath,amssymb]{revtex4} 

\usepackage{epsfig}
\usepackage{mathrsfs}
\usepackage{amsmath}
\usepackage{slashed}
\usepackage{subfigure}
\usepackage{placeins}
\usepackage{verbatim}

\newcommand{\ktsqav}{\langle k_{T}^{2}\rangle}
\newcommand{\ktsqavgT}{\langle k_{T}^{2}\rangle_{_{\rm WG}}} 
\newcommand{\ktsqavfT}{\langle k_{T}^{2}\rangle_{_{\rm S}}} 

\begin{document}
 
\title{Double Sivers effect asymmetries and their impact on transversity measurements at RHIC}
\author{Dani\"el Boer}
\email{D.Boer@rug.nl}
\affiliation{Theory Group, KVI, University of Groningen,
Zernikelaan 25, 9747 AA Groningen, The Netherlands}
\author{ Wilco J. den Dunnen}
\email{w.den.dunnen@vu.nl}
\affiliation{Department of Physics and Astronomy, 
Vrije Universiteit Amsterdam, 
De Boelelaan 1081, 1081 HV Amsterdam, The Netherlands}
\author{Aram Kotzinian}
\email{aram.kotzinian@cern.ch}
\affiliation{Dipartimento di Fisica Teorica, Universit\`a di Torino\\
         and \it INFN, Sezione di Torino, Via P. Giuria 1, I-10125 Torino, Italy\\
and \it Yerevan Physics Institute, 2 Alikhanyan Brothers ST., 375036 Yerevan, Armenia}

\date{\today}

\begin{abstract}
We study double transverse spin asymmetries in the Drell-Yan process at measured transverse momentum of the lepton pair.
Contrary to what a collinear factorization approach would suggest, a nonzero double transverse spin asymmetry
in the laboratory frame {\it a priori} does not imply nonzero transversity. TMD effects, such as the double Sivers effect, in principle form a 
background. Using the current knowledge of the relevant TMDs we estimate their contribution in the laboratory frame for Drell-Yan and $W$ production at RHIC and point out a cross-check asymmetry measurement to bound the TMD contributions. 
We also comment on the transverse momentum integrated asymmetries that only receive power suppressed background contributions.
\end{abstract}


\maketitle

\section{Introduction}

Transversity --the distribution of transversely polarized quarks inside a transversely 
polarized hadron-- was first discussed by Ralston and Soper \cite{Ralston:1979ys}, 
who suggested its measurement in the double polarized Drell-Yan process. Although this 
suggestion was made more than 30 years ago, this demanding double transverse
spin asymmetry measurement has not yet been performed. At present it is in the 
future physics program of BNL's Relativistic Heavy Ion Collider (RHIC) \cite{Bunce:2000uv}
and is also considered at GSI-FAIR, J-PARC and NICA. Currently RHIC is the only 
accelerator where polarized proton-proton collisions can be performed. Therefore, 
in this article we will focus on RHIC. 

Ralston and Soper considered the double transverse spin asymmetry ($A_{TT}$) 
integrated over the transverse momentum $q_T$ of the lepton pair (later also reconsidered in 
\cite{Artru:1989zv,Jaffe:1991ra,Cortes:1991ja}), and at measured $q_T$, in particular at $q_T=0$. 
Both cases will receive nonzero contributions from transversity. The asymmetry 
as a function of $q_T$ has been studied in Ref.\ \cite{Kawamura:2007ze} in a collinear 
Collins-Soper-Sterman resummation approach, showing it to be maximally of order 5\% and fairly 
flat in $q_T$ up to a few GeV. At measured $q_T$ there will however be background contributions
from transverse momentum dependence of partons, that have not yet been considered. We will study 
these contributions to the double transverse spin asymmetries at measured transverse momentum 
in both the Drell-Yan process and in $W$-boson production, where in the latter case one expects zero 
contribution from transversity within the Standard Model \cite{Bourrely:1994sc,Rykov:1999ru}.

At RHIC single transverse spin asymmetries ($A_N$) will be studied in $W$ production as well, with 
the goal of measuring the sign of the Sivers effect \cite{Kang:2009bp,Metz:2010xs}. This effect 
refers to the fact that the transverse momentum distribution of quarks 
inside a transversely polarized hadron can be asymmetric w.r.t.\ the spin 
direction \cite{Sivers:1989cc}. This spin-orbit coupling effect arises from initial and/or final state 
interactions and has a calculable process dependence (when factorization applies). The sign in Drell-Yan 
or $W$ production is predicted to be opposite to the one in semi-inclusive deep inelastic scattering (SIDIS), 
the process in which the Sivers effect asymmetry was first observed 
\cite{Airapetian:2004tw,Airapetian:2009ti,Alekseev:2010rw}. 
The Sivers effect may also generate background for the transversity double transverse spin 
asymmetries in the Drell-Yan process at measured $q_T$, through a Sivers effect in both incoming hadrons. This double Sivers effect will be investigated in this paper. Moreover, it can lead to a 
nonzero result in $W$ production, which could be mistaken for physics beyond the Standard Model, 
for instance from the complex mixing of $W$ bosons with a hypothetical 
$W'$ boson that appears in many extensions of the Standard Model \cite{Boer:2010mc}. 
We will study these aspects of the double Sivers effect contribution quantitatively in this paper. 

Besides the double Sivers effect we will include contributions from another transverse momentum dependent effect that was first discussed
by Ralston and Soper \cite{Ralston:1979ys}: it describes 
the distribution of longitudinally polarized quarks inside a transversely polarized hadron. 
Both effects are described by a transverse momentum dependent parton distribution (TMD): the Sivers effect by a TMD often denoted by $f_{1T}^\perp$ \cite{Boer:1997nt} and the other by $g_{1T}$ \cite{Tangerman:1994eh}. 
The latter function also appears in the analysis of the evolution equation of the twist-three function $g_T=g_1+g_2$ \cite{Bukhvostov:1983te,Bukhvostov:1984as,Ratcliffe:1985mp,Ji:1990br,Ali:1991em} and is in the literature sometimes referred to as one of the two ``worm gear'' functions \cite{Alekseev:2010dm}. We will adopt this convention and refer to $g_{1T}$ as the Worm Gear (WG) function, because the other worm gear function $h_{1L}^\perp$ will not be discussed here.

The expressions for the double Sivers and WG effects for Drell-Yan 
have been given in Ref.\ \cite{Boer:1999mm,Lu:2007ev,Arnold:2008kf}. 
Quantitatively, both effects have been studied
in Ref.\  \cite{Lu:2007ev} in polarized proton-antiproton Drell-Yan at GSI-FAIR, 
and the double WG effect in $W$ and $Z$ boson production at RHIC in Ref.\ \cite{Boer:2000er}. 
The latter study contains some
mistake in its Eq.\ (21) that will be corrected here, without altering the conclusions. 

As can be seen from the expressions in Ref.\ \cite{Boer:1999mm,Lu:2007ev,Arnold:2008kf}, one can consider a specific
frame, the so-called Collins-Soper frame, that in principle allows one to distinguish the double transverse spin 
asymmetries $A_{TT}(q_T)$ arising from transversity, the Sivers effect and the WG effect. 
Different angular dependences can be projected out allowing to single out a specific contribution.  Transversity leads to a spin asymmetry proportional to 
$\cos 2\phi_S^l$, where the azimuthal angle $\phi_{S}^l$ is measured between the spin plane and the lepton plane, whereas the other two effects lead to 
spin asymmetries independent of this lepton azimuthal angle. 
However, in the laboratory frame, as we will show in this paper, all three effects will contribute to the angular distribution $\cos 2\phi_S^l$. 
This is in contrast to what a collinear factorization approach would suggest, e.g., in the treatment as applied in Ref.\ \cite{Kawamura:2007ze}, the Sivers and WG contributions are absent from the start. In that approach the expression for $A_{TT}(q_T)$ in the lab frame will be only in terms of the transversity distribution (see also Ref.\ \cite{Tangerman:1994eh}), which might lead to wrong conclusions about transversity.

The lab frame is thus {\it a priori} not the right frame to extract the transversity distribution from the asymmetry $A_{TT}(q_T)$, however,
it is experimentally more `direct' to extract spin asymmetries in the lab frame. Analyzing the data in the lab frame might also be more accurate, because any additional uncertainties from the transformation to the CS frame will avoided.
Furthermore, an analysis in the CS frame and the lab frame could be cross-checked with each other if one knows the expected differences between the two frames. 
In $W$-boson production with a leptonic decay, on the other hand,  it is impossible to transform to the CS frame, because the neutrino will go unobserved rendering it impossible to determine the transverse momentum of the $W$ boson. This means that the double transverse spin asymmetries in $W$-boson production \emph{have} to be studied in the lab frame, where the transverse momentum dependent effects form a background for the new physics studies as proposed in \cite{Boer:2010mc}.
We think it is therefore important to know quantitatively 
the size of the spin asymmetries in the lab frame caused by partonic transverse momentum effects in the Drell-Yan process and $W$-boson production, which will be explored in the remainder of this paper.

\section{Drell-Yan cross section in TMD factorization}
\begin{figure}[h]
\parbox{\textwidth}{
\centering
\includegraphics[width=0.4\textwidth]{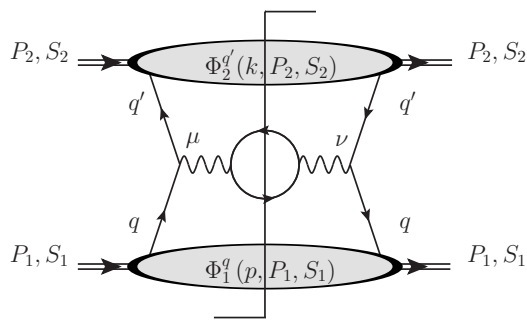}
}
\caption{Leading diagram in the Drell-Yan process.}\label{fig:DYgeneral}
\end{figure}

In both processes we have to deal with vector-boson production from hadron-hadron collisions, 
with a subsequent leptonic decay. 
The cross section for such a process has its leading contribution coming from the diagram in Fig.\ \ref{fig:DYgeneral}.
This contribution can be split into a lepton and hadron tensor connected by the appropriate vector-boson propagator.
The hadronic tensor can be expressed by
\begin{equation}\label{Wmunu}
W^{\mu\nu} =  \frac{1}{3} \sum_{q,q^\prime} \int d^4p d^4k \delta^4 (p+k-q)
\text{Tr} \Big[ \Phi^q(p,P_1,S_1) V^\nu_{qq^\prime} \Phi^{\bar{q}^\prime} (k,P_2,S_2) V^{\prime \mu}_{qq^\prime} \Big]
+ \left(1\leftrightarrow 2\right)
\end{equation}
in terms of the fully unintegrated quark correlator 
\begin{equation}
\Phi^q (p,P,S)_{ij} = \frac{1}{(2\pi)^4} \int d^4 \xi e^{ip\cdot \xi}
\langle P,S| \bar{q}_j (0) \mathcal{U}(0,\xi) q_i (\xi) |P,S\rangle
\end{equation}
and the quark--vector-boson vertices $V^\mu_{qq^\prime}$ and $V^{\prime \mu}_{qq^\prime}$, where the primed vertex has the complex conjugated coupling strength. The gauge link $\mathcal{U}(0,\xi)$ that renders the correlator gauge invariant is not specified at this stage. Writing the momenta 
in terms of lightcone and transverse components as $p=[p^-,p^+,\mathbf{p}_T]$, the delta function in the hadron tensor can be approximated by
\begin{equation}
\delta^4 (p+k-q) \approx \delta(p^+ - q^+) \delta(k^- - q^-)
\delta^2 (\mathbf{p}_T + \mathbf{k}_T - \mathbf{q}_T)
\end{equation}
which is accurate up to $p^- k^+/Q^2$ corrections, where $Q^2=q^2$ is the photon momentum squared. Keeping higher order corrections in the kinematics only gives very small corrections, see e.g.\ \cite{Anselmino:2005nn}.
This fixes $q^+ = p^+ \equiv x_1 P_1^+$ and $q^- = k^- \equiv x_2 P_2^-$ and allows us to write the hadron tensor as
\begin{equation}\label{HadronTensor}
W^{\mu\nu} = \frac{1}{3} \sum_{q,q^\prime} \int d^2\mathbf{p}_T d^2\mathbf{k}_T 
\delta^2(\mathbf{p}_T + \mathbf{k}_T - \mathbf{q}_T)
\text{Tr} \Big[ \Phi_1^q(x_1,\mathbf{p}_T) V^\nu_{qq^\prime} \Phi_2^{\bar{q}^\prime} (x_2,\mathbf{k}_T) V^{\prime \mu}_{qq^\prime} \Big] 
+ \left(1\leftrightarrow 2\right),
\end{equation}
in terms of the $p^-$-integrated quark correlator
\begin{equation}
\begin{aligned}
\Phi_1^q(x,\mathbf{p}_T)_{ij} &= \int dp^- \Phi^q(p,P_1,S_1)_{ij} \bigg|_{p^+=xP_1^+}
\\
&= \int \frac{d\xi^- d^2\mathbf{\xi}_T}{(2\pi)^3} e^{ip\cdot \xi}
\langle P_1,S_1| \bar{q}_j(0) \mathcal{U}(0,\xi)
q_i(\xi) |P_1,S_1 \rangle \bigg|_{\xi^+=0}.
\end{aligned}
\end{equation}
and a similar $\Phi_2$, which is integrated over $k^+$. The gauge link is process dependent, leading to a Sivers function in Drell-Yan that has the opposite sign compared to the one in SIDIS, cf.\  for instance Ref.\ \cite{Kang:2009bp} and references therein. 
In the double Sivers effect this sign is not relevant however.
The quark correlators can be parameterized in terms of transverse momentum dependent distribution functions by
\begin{equation}
\begin{aligned}
\Phi_1^q(x_1,\mathbf{p}_T) &= \frac{1}{2}\bigg\{ f_1^q (x_1,p_T) \slashed{n}_+
+ f_{1T}^{\perp q} (x_1,p_T) 
\frac{\epsilon_{\mu\nu\rho\sigma}\gamma^\mu n_+^\nu p_T^\rho S_{1T}^\sigma}{M_p}
+ g_{1T}^q(x_1,p_T)\frac{\mathbf{p}_T\cdot \mathbf{S}_{1T}}{M_p}\gamma_5\slashed{n}_+
+\cdots  \bigg\},
\\
\Phi_2^q(x_2,\mathbf{k}_T) &= \frac{1}{2} \bigg\{ f_1^q (x_2,k_T) \slashed{n}_-
+ f_{1T}^{\perp q} (x_2,k_T) 
\frac{\epsilon_{\mu\nu\rho\sigma}\gamma^\mu n_-^\nu k_T^\rho S_{2T}^\sigma}{M_p}
+ g_{1T}^q(x_2,k_T)\frac{\mathbf{k}_T\cdot \mathbf{S}_{2T}}{M_p}\gamma_5\slashed{n}_-
+\cdots  \bigg\},
\end{aligned}
\end{equation}
in which we only wrote the relevant distribution functions for the transverse spin asymmetries to leading order in $M_p/P_1^+$ and $M_p/P_2^-$.
For further details we refer to, e.g., Ref.\ \cite{Boer:1999mm}.

We will define a spin flip symmetric and antisymmetric cross section by
\begin{equation}\label{CSinWL}
\begin{aligned}
d\sigma^S	&\equiv \frac{1}{4}\left(
		d\sigma^{\uparrow\uparrow} + d\sigma^{\uparrow\downarrow} 
		+ d\sigma^{\downarrow\uparrow} + 
		d\sigma^{\downarrow\downarrow} \right)
		=\frac{1}{2s} W_S^{\mu\nu} D_{\mu\rho} 
		D^*_{\nu\sigma}L^{\rho\sigma} dP,\\
d\sigma^A	&\equiv \frac{1}{4}\left(
		d\sigma^{\uparrow\uparrow} - d\sigma^{\uparrow\downarrow} 
		- d\sigma^{\downarrow\uparrow} 
		+ d\sigma^{\downarrow\downarrow} \right)
		=\frac{1}{2s} W_A^{\mu\nu} D_{\mu\rho}
		D^*_{\nu\sigma}L^{\rho\sigma} dP,
\end{aligned}
\end{equation}
where $W^{\mu\nu}_{S,A}$ is the hadron tensor symmetrized or antisymmetrized with respect to the proton spins, $D_{\mu\rho}$ is the vector-boson propagator,
$L^{\rho\sigma}$ is the lepton tensor
\begin{equation}
L^{\rho\sigma} = \text{Tr}\left[V_l^\rho \slashed{\bar{l}} V_l^\sigma \slashed{l} \right]
\end{equation}
in terms of the lepton-vector-boson vertex $V_l^\rho$ and, finally, $dP$ is the phase space element
\begin{equation}\label{dPdef}
\begin{aligned}
dP 	= (2\pi)^4 \frac{d^3\vec{l}}{(2\pi)^3 2 E_l} 
		\frac{d^3\vec{\bar{l}}}{(2\pi)^3 2 E_{\bar{l}}}.
\end{aligned}
\end{equation}

\section{Distribution functions}

The distribution function $f_1^q(x,k_T)$ describes the probability of finding a quark $q$ with lightcone momentum fraction $x$ and transverse momentum with length $k_T \equiv |\mathbf{k}_T|$. As often done, 
for our phenomenological studies we will assume factorization between $k_T$ and $x$ dependence 
and assume a Gaussian dependence on $k_T$, i.e.
\begin{equation}
f_1^q(x,k_T) = \frac{1}{\pi \ktsqav} e^{-k_T^2/\ktsqav} f_1^q(x).
\end{equation}
Such a Gaussian dependence has been shown to work very well \cite{Schweitzer:2010tt}. We will use the value of the width,
\begin{equation}
\ktsqav = 0.25\ \text{GeV}^2,
\end{equation}
found by \cite{Anselmino:2005nn} based on the Cahn effect in unpolarized SIDIS. Although this value may differ from the  
$\ktsqav$ in Drell-Yan, the deviation is not expected to matter for our purposes and to fall within the error in the estimates we will consider.

The Sivers distribution function describes the correlation between the partonic transverse momentum and proton spin direction. The probability of finding a quark $q$ with transverse momentum $\mathbf{k}_T$ inside a transversely polarized proton is given by
\begin{equation}
\mathcal{P}^q(x,\mathbf{k}_T) = f_1^q(x,k_T) 
	+ \sin(\phi_{\mathbf{k}_T} - \phi_{\mathbf{S}_T})\frac{|\mathbf{k}_T||\mathbf{S}_T|}{M_p}
	f_{1T}^{\perp q} (x,k_T).
\end{equation}
In SIDIS there are clear experimental observations of the asymmetries that would arise from the Sivers effect, offering strong support for the latter effect.
Within that picture the current experimental data allows for a determination of the Sivers function for both the $u$ and $d$ quarks \emph{and} anti-quarks. In the recent extraction obtained by \cite{Anselmino:2008sga}, the Sivers function for SIDIS is parameterized as
\begin{equation}
f_{1T}^{\perp q}(x,k_T)	= -\mathcal{N}_q(x) h^\prime(k_T) f_1^q(x,k_T) 
\end{equation}
with 
\begin{equation}
\begin{aligned}
h^\prime (k_T) 		&= \sqrt{2e} \frac{M_p}{M_1} e^{-k_T^2/M_1^2},\\
\mathcal{N}_q(x) 	&=  N_q x^{\alpha_q}(1-x)^{\beta_q} 
			\frac{(\alpha_q + \beta_q)^{\alpha_q + \beta_q}}
			{\alpha_q^{\alpha_q} \beta_q^{\beta_q}}.
\end{aligned}
\end{equation}
The numerical values found in the extraction are $M_1^2= 0.34 \text{ GeV}^2$ for the flavor independent width of the distribution and the numbers in table \ref{Siversparams} for the parameters in the flavor dependent function that describes the $x$ dependence. The current knowledge of the Sivers function at small $x$ is limited, but the single spin asymmetry measurements at RHIC will certainly improve this. For the moment, we take what is known until a better determination will be available. Taking into account the error bars in \cite{Anselmino:2008sga} we come to the rough estimate that the overall effect, as will be calculated in Section \ref{DY} and \ref{Wproduction}, can be maximally enhanced by an order of magnitude. 
As said before, the sign of the Sivers function for Drell-Yan is supposed to be opposite to the one for SIDIS, however in the \emph{double} Sivers effect this has no influence.

\begin{table}[!ht]
\centering
\begin{tabular}{l|cccc}
		&$u$	&$\bar{u}$	&$d$	&$\bar{d}$\\	
\hline
$\alpha_q$	&0.73	&0.79		&1.08	&0.79\\
$\beta_q$	&3.46	&3.46		&3.46	&3.46\\
$N_q$		&0.35	&0.04		&-0.9	&-0.4\\
\end{tabular}
\caption{Numerical values for the parameters in the Sivers function from \cite{Anselmino:2008sga}.}
\label{Siversparams}
\end{table}

The Worm Gear distribution $g_{1T}^q(x,k_T)$ describes the \emph{longitudinal} polarization of quarks with transverse momentum 
$\mathbf{k}_T$, inside a \emph{transversely} polarized proton. A determination of this distribution based on fits of experimental data is not available. Data on double transverse spin asymmetries $A_{LT}$ that receive contributions from the WG effect has become 
available only very recently \cite{Parsamyan:2010se,XiaodongJiangINT2010}. The recent measurements on $^3$He indicate that $g_{1T}$ for the up-quark is not small \cite{XiaodongJiangINT2010}.

Here we will employ a model for this WG function.
Both the bag model \cite{Avakian:2010br} and the spectator model \cite{Jakob:1997wg} agree quite well with a Gaussian approximation of the transverse momentum dependence for not too large values of the transverse momentum. 
We will therefore use the Gaussian Ansatz, which allows us to express the transverse momentum dependent distribution as
\begin{equation}
g_{1T}^q(x_,k_T) = \frac{2 M_p^2}{\pi\ktsqavgT^2} e^{-k_T^2/\ktsqavgT} 			g_{1T}^{q(1)} (x),
\end{equation}
in terms of its first transverse moment $g_{1T}^{q(1)} (x)$, which is defined as
\begin{equation}\label{g1T1def}
g_{1T}^{q(1)}(x) \equiv \int d^2 k_T \frac{k_T^2}{2M_p^2} g_{1T}^q(x,k_T).
\end{equation}
For the width we will take a value in accordance with the bag model
\begin{equation}
\ktsqavgT=0.71\ktsqav.
\end{equation}
For the first moment, we will 
use a Wandzura-Wilczek (WW) type approximation \cite{Wandzura:1977qf,Tangerman:1994bb,Kotzinian:1995cz} to express it in terms of the known helicity distribution $g_1(x)$ by
\begin{equation}
g_{1T}^{q(1)}(x) \approx x \int_x^1 dy \frac{g_1^q(y)}{y}.
\end{equation}
For numerical estimations of this function the DSSV helicity distribution \cite{deFlorian:2008mr} will be used. 
Deviations from the WW approximation can be considered \cite{Accardi:2009au}, 
but the WW distribution is in fair agreement with the bag model, the spectator model, the light cone constituent quark model \cite{Pasquini:2008ax} and the light cone quark-diquark model \cite{Zhu:2011zz}. 
Furthermore, a recent determination of target transverse spin asymmetries in SIDIS \cite{Parsamyan:2007ju} is consistent with the theoretical prediction based on the WW type approximation of \cite{Kotzinian:2006dw}.
With all these ingredients the lowest Mellin moment $g_{1T}^{q(0,1)}\equiv \int dx g_{1T}^{q(1)}(x)$ of the first transverse moment $g_{1T}^{q(1)}(x)$ can be calculated. We find $g_{1T}^{u(0,1)}=0.091$ and $g_{1T}^{d(0,1)}= -0.026$. This is in excellent agreement with the evaluation on the lattice (at the scale 1.6 GeV): $g_{1T}^{u(0,1)}=0.1055(66)$ and $g_{1T}^{d(0,1)}=-0.0235(38)$ from Ref.\ \cite{Hagler:2009mb}. All this gives us confidence that the estimates below are sufficiently realistic. 

\section{Spin asymmetries in the Drell-Yan process}\label{DY}

In the Drell-Yan process the virtual photon produces a lepton and anti-lepton, both of which can be detected. 
This allows a full determination of all the kinematic variables, such that one can transform to the so-called Collins-Soper frame \cite{Collins:1977iv}.
In that frame a correlation between the lepton transverse momentum and the proton transverse spin direction will come solely from the transversity distribution, which makes this process very suitable for an extraction of this distribution function from $A_{TT}(q_T)$.
If one just analyzes the correlation between the lepton angle and the proton spin direction in the lab frame, there can be a residual asymmetry coming from double Sivers and WG effects for the following reason. 
The Sivers and WG function both cause the photon transverse momentum and the proton spins to be correlated. When the virtual photon decays, the decay products are more inclined to move in the direction of the parent particle which, in turn, causes the direction of the decay products to be \emph{also} correlated with the proton spin directions, albeit diluted.

In order to estimate the error that one would possibly make in the extraction of the transversity distributions from $A_{TT}(q_T)$ by not going to the Collins-Soper frame, we will calculate the double transverse spin asymmetries in the lab frame coming from the Sivers and WG functions.

In the following analysis we will work towards an asymmetry differential in the photon's momentum squared $Q^2$, transverse momentum length $q_T$, and rapidity $Y\equiv \frac{1}{2}\log q^+/q^-$.
The other kinematic variables, which will be integrated over, are $\phi_q$, which is the azimuthal angle of $\mathbf{q}_T$, and $y$, which is defined as $y\equiv l^-/q^-$.
The final kinematic variable is the direction of the lepton transverse momentum $\phi_l$ in the lab frame, which in the end will be integrated over with particular weights to select out the different contributions to the spin asymmetries.

\begin{figure}[htb]
\subfigure[\ $Q=5$ GeV and $Y=0$]
{\includegraphics[height=3cm]{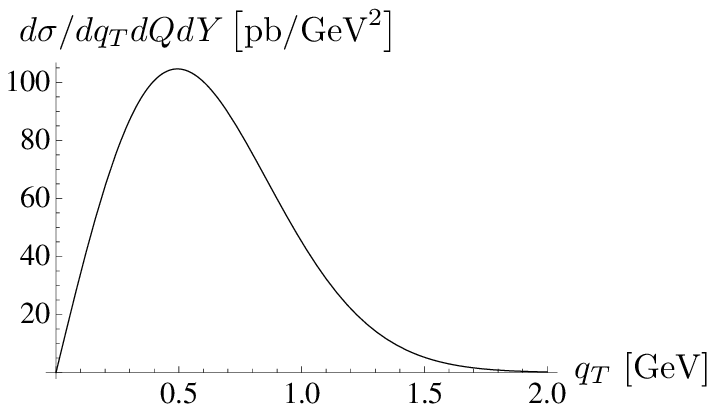}}
\subfigure[\ $q_T=1$ GeV and $Y=0$]
{\includegraphics[height=3cm]{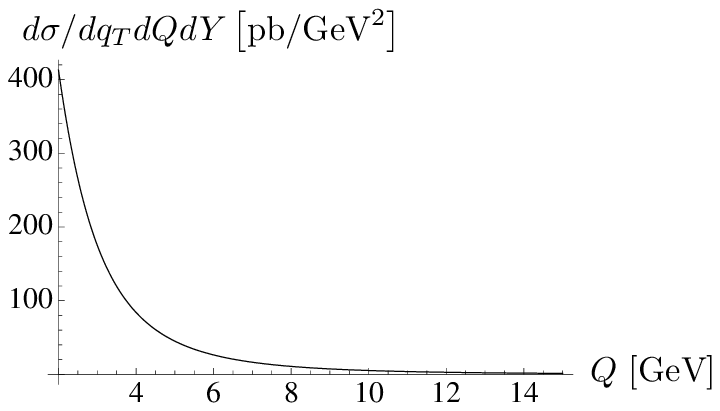}}
\subfigure[\ $Q=5$ GeV and $q_T=1$ GeV]
{\includegraphics[height=3cm]{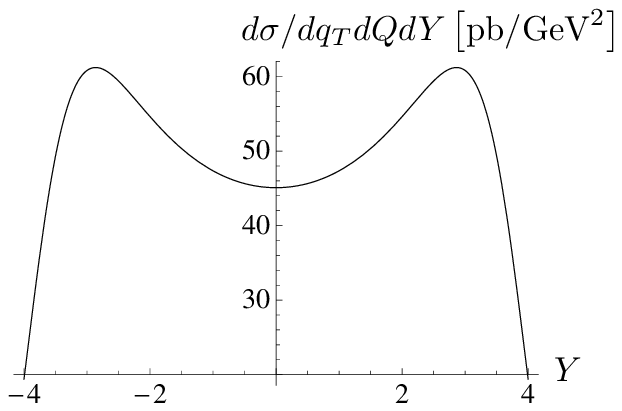}}
\caption{Differential cross section for the Drell-Yan process at RHIC energy $\sqrt{s}=500$ GeV.}\label{fig:SigmaDY}
\end{figure}

Starting from Eq.\ \eqref{CSinWL}, we can express the cross section as
\begin{equation}
\frac{d\sigma}{dQ dq_T d\phi_l dY}
	= \int dy d\phi_q
	\frac{q_T}{8(2\pi)^2 s Q^3}
	\Bigg( 1 + \frac{q_T\cos(\phi_l - \phi_q)}{\sqrt{Q^2\frac{1-y}{y}
	- q_T^2\sin^2(\phi_l - \phi_q)}} \Bigg)
	W^{\mu\nu}L_{\mu\nu},\\
\end{equation}
by using the fact that effectively the photon propagator $D_{\mu\nu}=-ig_{\mu\nu}/Q^2 $ and the phase space element in these lab frame coordinates is 
\begin{equation}
dP	= \frac{Q q_T}{4 (2\pi)^2} dq_T d\phi_q d\phi_l dQ dY dy
	\Bigg(1 + \frac{q_T \cos(\phi_l - \phi_q)}{\sqrt{Q^2\frac{1-y}{y}
	-q_T^2\sin^2(\phi_l - \phi_q)}} \Bigg).
\end{equation}
The vertices in the lepton and hadron tensor are, for the Drell-Yan process, given by
\begin{equation}
\begin{aligned}
V_{qq^\prime}^\nu 	&= i e_q e \gamma^\nu \delta_{qq^\prime},\\
V_l^\rho		&= -i e \gamma^\rho.
\end{aligned}
\end{equation}
Furthermore, in the expression for the lepton tensor we need the lepton ($l$) and anti-lepton ($\bar{l}$) momentum 4-vectors, which can be specified in terms of lightcone and transverse components, 
in the lab frame by
\begin{equation}
\begin{aligned}
l 	&= \Bigg[
	\frac{1}{\sqrt{2}}\sqrt{\frac{y}{1-y}} e^{-Y}\sqrt{(1-y)l_T^2+y\bar{l}_T^2}, 
	\frac{\frac{1}{\sqrt{2}} \sqrt{\frac{1-y}{y}}l_T^2
	e^Y}{\sqrt{(1-y)l_T^2+y\bar{l}_T^2}},
	\mathbf{l}_T  \Bigg],
\\
\bar{l}	&= \Bigg[ 
	\frac{1}{\sqrt{2}}\sqrt{\frac{1-y}{y}} e^{-Y}\sqrt{(1-y)l_T^2+y\bar{l}_T^2},
	\frac{\frac{1}{\sqrt{2}}\sqrt{\frac{y}{1-y}}\bar{l}_T^2e^Y}
	{\sqrt{(1-y)l_T^2+y\bar{l}_T^2}},
	\mathbf{\bar{l}}_T  \Bigg],
\end{aligned}
\end{equation}
where 
\begin{equation}
\bar{l}_T = \sqrt{q_T^2 + l_T^2 - 2 l_T q_T \cos(\phi_l - \phi_q)}
\end{equation}
and
\begin{equation}
l_T = q_T y \cos(\phi_l - \phi_q) + \sqrt{Q^2 y (1-y) 
- q_T^2 y^2 \sin^2(\phi_l - \phi_q)}.
\end{equation}
The lightcone momentum fractions are in terms of the lab-frame coordinates given by
\begin{equation}
x_{1,2}	= e^{\pm Y} \sqrt{\frac{Q^2+q_T^2}{s}}.
\end{equation}
Having all those ingredients the lepton and hadron tensor can be calculated. In the hadron tensor the $\mathbf{k}_T$ and $\mathbf{p}_T$ integrals are performed. After contracting the lepton and hadron tensor using Eq.\ \eqref{CSinWL}, the resulting expression for the cross section is integrated over $y$ and expanded in powers of $q_T/Q$ except for the Gaussian in the distributions, which delivers the high $q_T$ suppression, and the expression for $x_{1,2}$ in the distribution functions. 
This expansion allows us to perform the $\phi_q$ integration analytically. 
After having done the $\phi_q$ integration we obtain the following approximate expression for the symmetric cross section,
\begin{equation}
\frac{d\sigma^S}{dq_T dQ d\phi_l dY} = \sum_q
\frac{4 \alpha^2 e_q^2 q_T}{9 \ktsqav Q s} 
e^{-{q_T^2}/{2 \ktsqav}} F_1^q(x_1,x_2),
\end{equation}
which is accurate up to leading order in $\mathcal{O}(q_T/Q)$, furthermore we have defined
\begin{equation}
F_1^q (x_1,x_2) \equiv f_1^q(x_1) f_1^{\bar q}(x_2) +
f_1^q(x_2) f_1^{\bar q}(x_1).
\end{equation}
This cross section integrated over $\phi_l$ is plotted as function of the three remaining variables in Fig.\ \ref{fig:SigmaDY}.

For the antisymmetric cross section we find the expression,
\begin{equation}
\begin{aligned}
\frac{d\sigma^A}{dq_T dQ d\phi_l dY} =
\sum_q \frac{\alpha^2 e_q^2 |\mathbf{S}_T|^2 q_T}{9 M_p^2 Q s}
\Bigg\{ &e^{- q_T^2 / 2\ktsqavfT} 
\bigg[ 1 - \frac{q_T^2}{2\ktsqavfT} + \frac{q_T^4}{16 Q^2 \ktsqavfT}
\cos 2 \phi_{S}^l \bigg]
F_{1T}^{\perp q} (x_1,x_2)\\
& + e^{- q_T^2/2 \ktsqavgT} \bigg[ -1 + \frac{q_T^2}{2\ktsqavgT} + \frac{q_T^4}{16 Q^2 \ktsqavgT} \cos 2 \phi_{S}^l \bigg] 
G_{1T}^q(x_1,x_2)\Bigg\},
\end{aligned}
\end{equation}
in which we kept leading order terms in $q_T/Q$ only, which is for the $\phi_l$ independent part $\mathcal{O}(q_T/Q)$ and for the $\cos2\phi_S^l$ dependent part $\mathcal{O}(q_T^3/Q^3)$. Furthermore, $\phi^l_S\equiv\phi_S-\phi_l$ is the angle between the spin plane and the lepton transverse momentum and
\begin{equation}
\begin{aligned}
F_{1T}^{\perp q}(x_1,x_2) &\equiv f_{1T}^{\perp q}(x_1) f_{1T}^{\perp \bar{q}}(x_2) +
f_{1T}^{\perp q}(x_2) f_{1T}^{\perp \bar{q}}(x_1),
\\
G_{1T}^q(x_1,x_2) &\equiv g_{1T}^q(x_1) g_{1T}^{\bar q}(x_2) +
g_{1T}^q(x_2) g_{1T}^{\bar q}(x_1), 
\end{aligned}
\end{equation}
in terms of $f_{1T}^{\perp q}(x)$ and $g_{1T}^q(x)$, which are defined through the relation
\begin{equation}
\begin{aligned}
f_{1T}^{\perp q}(x,k_T)	&= \frac{1}{\pi \ktsqavfT} e^{-k_T^2/\ktsqavfT} 
			f_{1T}^{\perp q}(x),\\
g_{1T}^q(x,k_T) 	&= \frac{1}{\pi \ktsqavgT} e^{-k_T^2/\ktsqavgT} 
			g_{1T}^q(x),\\
\end{aligned}
\end{equation}
in which
\begin{equation}\label{ktsqavSdef}
\ktsqavfT \equiv \frac{\ktsqav M_1^2}{\ktsqav + M_1^2}.
\end{equation}
We will define the three spin asymmetries
\begin{equation}\label{ATTdef}
\begin{aligned}
A_{TT}^0(q_T)	&\equiv \frac{\int_0^{2\pi}d\phi_l d\sigma^A}
		{\int_0^{2\pi}d\phi_l d\sigma^S},\\
A_{TT}^C(q_T) 	&\equiv \frac{\left(\int_{-\pi/4}^{\pi/4} -
  		\int_{\pi/4}^{3\pi/4} + \int_{3\pi/4}^{5\pi/4} -
  		\int_{5\pi/4}^{7\pi/4}\right)d\phi_l d\sigma^A
  		}{\int_0^{2\pi}d\phi_l d\sigma^S},\\
A_{TT}^S(q_T) 	&\equiv \frac{\left(\int_{0}^{\pi/2} -
		\int_{\pi/2}^{\pi} + \int_{\pi}^{3\pi/2} -
		\int_{3\pi/2}^{2\pi}\right)d\phi_l d\sigma^A 
		}{\int_0^{2\pi}d\phi_l d\sigma^S},
\end{aligned}
\end{equation}
which select out the $\phi_S^l$ independent, cosine modulated and sine modulated terms, respectively. The latter,  $A_{TT}^S(q_T)$, will be zero in this case, but not in $W$ production (cf.\ next section) or $\gamma$-$Z$ interference (not considered here). Both the $A_{TT}^C(q_T)$ asymmetry, to which transversity contributes, 
and the $A_{TT}^0(q_T)$ asymmetry receive a nonzero contribution from the double Sivers and WG effects and can be written as
\begin{multline}
A_{TT}^0(q_T) = 
\frac{|\mathbf{S}_T|^2 \ktsqav}{4 M_p^2} 
\Bigg\{e^{- \frac{q_T^2}{2M_1^2}} \left(1-\frac{q_T^2}{2\ktsqavfT}\right)
\frac{\sum_q e_q^2 F_{1T}^{\perp q}(x_1,x_2)}{\sum_q e_q^2 F_1^q(x_1,x_2)}\\
- e^{- q_T^2 \frac{\ktsqav-\ktsqavgT}{2 \ktsqav \ktsqavgT}} \left(1-\frac{q_T^2}{2\ktsqavgT} \right)
\frac{\sum_q e_q^2 G_{1T}^q(x_1,x_2)}{\sum_q e_q^2 F_1^q(x_1,x_2)} \Bigg\}
\end{multline}
and
\begin{equation}
A_{TT}^C(q_T) =
\frac{|\mathbf{S}_T|^2 \ktsqav q_T^2}{32 \pi M_p^2 Q^2} 
\left\{e^{- \frac{q_T^2}{2M_1^2}} \frac{q_T^2}{\ktsqavfT}
\frac{\sum_q e_q^2 F_{1T}^{\perp q}(x_1,x_2)}{\sum_q e_q^2 F_1^q(x_1,x_2)}
+ e^{- q_T^2 \frac{\ktsqav-\ktsqavgT}{2 \ktsqav \ktsqavgT}} \frac{q_T^2}{\ktsqavgT}
\frac{\sum_q e_q^2 G_{1T}^q(x_1,x_2)}{\sum_q e_q^2 F_1^q(x_1,x_2)} \right\}.
\end{equation}
We note that the bound on the $\cos 2\phi_{S}^l$ double transverse spin asymmetry as a function of $q_T$ from transversity was
estimated, within a collinear Collins-Soper-Sterman resummation approach \cite{Kawamura:2007ze}, to be maximally of order 5\% and fairly flat in $q_T$ up to a few GeV at RHIC at a center of mass energy of 500 GeV. 
The first extraction of the quark transversity distribution $h_1^{q}$ \cite{Anselmino:2007fs,Anselmino:2008jk}, however, indicates it to be about half its maximally allowed value at $Q^2 \sim 2$ GeV$^2$. Therefore, if this also applies to the antiquark $h_1^{\bar q}$, an asymmetry of 1\% or less should be expected at RHIC. 

\begin{figure}[htb]
\subfigure[\ $Q=5$ GeV and $Y=0$]
{\includegraphics[height=2.6cm]{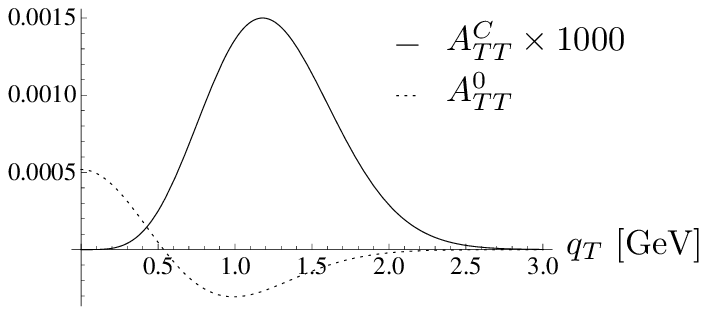}}
\subfigure[\ $q_T=1$ GeV and $Y=0$]
{\includegraphics[height=2.6cm]{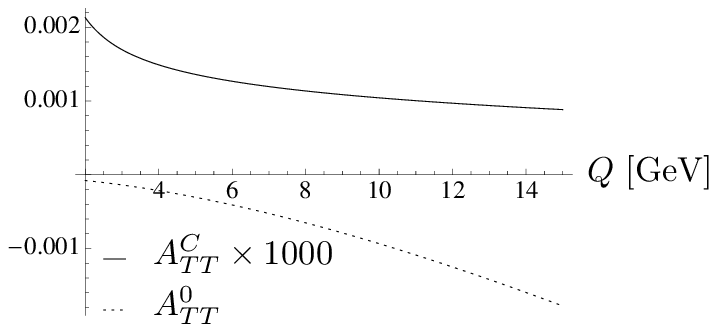}}
\subfigure[\ $Q=5$ GeV and $q_T=1$ GeV]
{\includegraphics[height=2.6cm]{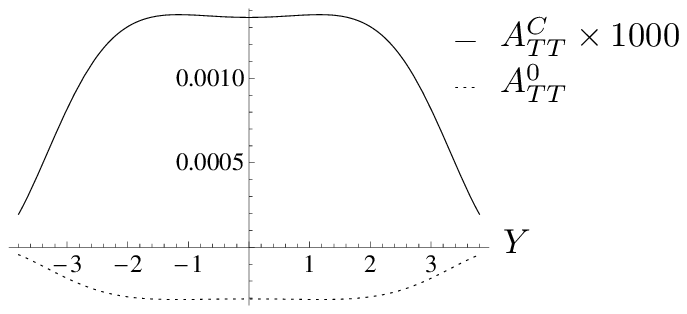}}
\caption{Contribution to $A_{TT}(q_T)$ in the Drell-Yan process from the double Sivers effect at RHIC energy $\sqrt{s}=500$ GeV.}\label{ATTDY}
\end{figure}

Asymmetries that are below the permille level in the entire kinematic range of interest will generally not be shown. They will be below the detection limit at RHIC, which will be mainly restricted by systematic errors. For the case of Drell-Yan this will only leave the Sivers effect contribution to the asymmetry $A_{TT}^0(q_T)$, displayed in Fig.\ \ref{ATTDY} as function of $q_T$, $Q$ and $Y$. 
In the plot we also included, albeit completely negligible, the Sivers effect contribution to $A_{TT}^C(q_T)$, just in case the Sivers function at these values of $x$ and $Q$ turns out to be much larger. 

The $A_{TT}^0(q_T)$ asymmetry reaches up to the percent level, but only for large $Q^2$ outside the range of interest. In the standard Drell-Yan range between the $J/\psi$ and $\Upsilon$, the asymmetry is on the permille level for the double Sivers effect and far below that level for the double WG effect.  

The $A_{TT}^C(q_T)$ asymmetry receives 
a contribution from the double Sivers effect at a level of $10^{-6}$ and from the $g_{1T}$ function a contribution at a level of $10^{-8}$. At small $Q$ the asymmetry is small due to the smallness of the Sivers function with respect to the unpolarized distribution at low values of $x$, whereas at higher values of $Q$ the $q_T^2/Q^2$ suppression becomes important.
One way or the other, these magnitudes are far below the detection limit at RHIC, even if one takes into account a possible enhancement of the effect by an order of magnitude due to the uncertainty in the used parametrization of the Sivers function.
Therefore, the TMD effects will not spoil a determination of the transversity distribution if those are determined from $A_{TT}^C(q_T)$ in the lab frame instead of in the Collins-Soper frame. As a cross-check, to assure that TMD effects are small, one could verify that the $A_{TT}^0(q_T)$ asymmetry is small. 
The $A_{TT}^C(q_T)$ asymmetry is bounded by the larger $A_{TT}^0(q_T)$ asymmetry due to the $q_T^2/Q^2$ suppression, irrespective of any assumptions on the Sivers function or the Worm-Gear distribution.
We want to note that, considering asymmetries of this size, higher twist effects could become important. 
In case of incomplete averaging over the azimuthal angle, the $\phi_l$ independent asymmetry $A_{TT}^0(q_T)$ may form a background for a determination of the $\phi_l$ dependent $A_{TT}^C(q_T)$, but given its magnitude this should also not pose a problem. 

The $q_T$-integrated asymmetries have also been calculated and were found to be a factor two smaller for $A_{TT}^C$ 
and a factor 1000 smaller for $A_{TT}^0$. This agrees with the expectation that such effects are (at least) 
${\cal O}(M_p^2/Q^2)$ power suppressed in this case. For completeness, we mention that the maximal 
$q_T$-integrated $A_{TT}^C$ asymmetry from transversity is estimated to be at the few percent level at RHIC at a center of mass energy of 200 and 500 GeV \cite{Martin:1996mg,Martin:1997rz}.

\FloatBarrier

\section{Spin asymmetries in $W$ boson production}\label{Wproduction}

In $W$-boson production one expects zero 
contribution from transversity within the Standard Model \cite{Bourrely:1994sc,Rykov:1999ru}. 
This strong prediction offers the possibility to study contributions from a hypothetical $W'$ boson that appears in many extensions of the Standard Model. It is at least ten times heavier 
than the $W$ boson and is expected to produce double transverse spin asymmetries at RHIC 
only through its mixing with the $W$ boson, which gives the latter a small righthanded coupling to 
fermions. In Ref.\ \cite{Boer:2010mc} this was studied quantitatively and it was shown that RHIC 
may be able to produce competitive bounds if the original design goals are met and if $h_1^{\bar q}/f_1^{\bar q}$ is not very much 
smaller than $h_1^{q}/f_1^{q}$. It was also shown 
that all other Standard Model background was well below the permille level, but for the background from the 
double Sivers and WG effect this was not yet quantitatively estimated.
In the leptonic decay of a $W$ boson the neutrino will go unobserved, which renders it impossible to determine the Collins-Soper frame and remove the background.
Therefore, we will estimate quantitatively the sizes of these Standard Model effects to see whether it would jeopardize any measurement of this $W-W^\prime$ mixing using spin asymmetries.

In Ref.\ \cite{Boer:2010mc} the TMD background was dismissed on the basis of 
a dimensional counting argument. If the single Sivers effect is a 10\% effect, a double Sivers effect asymmetry 
in $W$ production would be on the percent level. However, as the Sivers asymmetry is 
an azimuthal asymmetry, the lack of the knowledge of the $W$ momentum prevents 
reconstruction of the asymmetry in $W$ production directly. Instead, a lepton asymmetry
can be measured (cf.\ \cite{Nadolsky:2003ga}), which has a reduced magnitude. Naively one would expect from a dimensional 
analysis a large suppression of the size $q_T^2/l_T^2$, where $q_T$ denotes the gauge boson 
transverse momentum and $l_T$ the lepton transverse momentum. The reason being that the 
asymmetry should vanish in the limit $q_T \to 0$ and the only compensating scale is $l_T$. This yields
an asymmetry well below the permille level. A similar argument would suggest the single spin 
asymmetry $A_N(q_T)$ in $W$ production arising from the Sivers effect to be $q_T/l_T$ suppressed, leading 
to an asymmetry below the percent level. However,
the Sivers effect in $A_N(q_T)$ in $W$ production has recently been studied theoretically \cite{Kang:2009bp} 
and a large asymmetry (of order 10\%) was predicted. Moreover, in Ref.\ \cite{Metz:2010xs} the lepton asymmetry $A_N(l_T)$
was evaluated numerically, which has a reduced magnitude, but still is around 3\% for $W^+$ production. 
This is larger than expected from the dimensional argument and is likely because   
near resonance the width of the $W$ boson becomes an important scale. 
The suppression can therefore be only as small as $q_T/\Gamma_W$. 

Similarly, a double Sivers effect contribution to $A_{TT}(q_T)$ in $W$ production is expected to be on the percent level and a factor $q_T^2/\Gamma_W^2$ smaller for the lepton asymmetry $A_{TT}(l_T)$ 
near resonance. When integrated over $l_T$ instead, one can 
expect the asymmetry to be suppressed by a factor of $q_T^2/M_W^2$, which implies an asymmetry well below the permille level. 
Below we confirm these insights in an explicit calculation.  

Starting from Eq.\ \eqref{CSinWL}, the spin symmetric and antisymmetric cross section for the production of a charged lepton from a $W^-$ boson decay can be expressed as
\begin{equation}
\frac{d\sigma^{S,A}}{dl_TdY_ld\phi_l} 
		=  \frac{l_T}{8s(2\pi)^2}
		\int dY_{\bar{l}} \int d^2\mathbf{q}_T 
		W_{S,A}^{\mu\nu} D_{\mu\rho} D^*_{\nu\sigma}L^{\rho\sigma},
\end{equation}
where we used $\mathbf{q}_T =  \mathbf{l}_T + \mathbf{\bar{l}}_T$ to write
$d^2\mathbf{\bar{l}}_T = d^2\mathbf{q}_T$. For $W^-$ production the quark and lepton vertices are
\begin{equation}
\begin{aligned}
V_{qq^\prime}^\mu	&=\frac{ig}{\sqrt{2}} (V_{CKM}^{ud})^* \gamma^\mu 				\mathcal{P}_L \delta_{uq^\prime}\delta_{dq},\\
V_l^\rho		&= \frac{ig}{\sqrt{2}} \gamma^\rho \mathcal{P}_L.
\end{aligned}
\end{equation}
The charged lepton ($l$) and neutrino ($\bar{l}$) momentum 4-vectors can, in the lab frame, be expressed by
\begin{equation}
l 	= \left[ \frac{l_T}{\sqrt{2}}e^{-Y_l}, \frac{l_T}{\sqrt{2}} e^{Y_l},
	\mathbf{l}_T \right], \qquad
\bar{l}	= \left[ \frac{\bar{l}_T}{\sqrt{2}}e^{-Y_{\bar{l}}},
	\frac{\bar{l}_T}{\sqrt{2}} e^{Y_{\bar{l}}}, \mathbf{q}_T-\mathbf{l}_T \right],
\end{equation}
in terms of the neutrino rapidity $Y_{\bar l}$ and charged lepton transverse momentum $\mathbf{l}_T$ and rapidity $Y_l$. Having those ingredients, we calculate the contraction of the lepton and hadron tensor as in Eq.\ \eqref{CSinWL} to get the cross section. The lightcone momentum fractions can be expressed in terms of $l_T$ and through a power expansion in $q_T$ as
\begin{equation}
\begin{aligned}
x_1 &= \frac{l_T}{\sqrt{s}}\left(e^{Y_l}+e^{Y_{\bar{l}}}\right)
	- \frac{q_T}{\sqrt{s}}e^{Y_{\bar{l}}}\cos(\phi_l-\phi_q)
	+ \mathcal{O}\left(\frac{q_T^2}{s}\right),\\
x_2 &= \frac{l_T}{\sqrt{s}}\left(e^{-Y_l}+e^{-Y_{\bar{l}}}\right)
	- \frac{q_T}{\sqrt{s}}e^{-Y_{\bar{l}}}\cos(\phi_l-\phi_q)
	+ \mathcal{O}\left(\frac{q_T^2}{s}\right).\\
\end{aligned}
\end{equation}
As we are working at leading twist only, we can drop the non-leading terms in this expression as well. The advantage is that there will not be any $\mathbf{q}_T$ dependence in the distribution functions, which allows us to perform the $\mathbf{q}_T$ integration in the cross section analytically. After having done the $\mathbf{q}_{T}$ integration analytically, we expand in the cross section in parton transverse momentum up to order $k_T^2$ and $p_T^2$. The integration with respect to $\mathbf{k}_T$ and $\mathbf{p}_T$ coming from the expression for the hadron tensor in Eq.\ \eqref{HadronTensor} can now be done, which results in an expression in terms of $g_{1T}^{(1)}(x)$, defined in Eq.\ \eqref{g1T1def}, and $f_{1T}^{\perp(1)}(x)$, which is likewise defined as
\begin{equation}
f_{1T}^{q(1)}(x) \equiv 
		\int d^2\mathbf{k}_T \frac{k_T^2}{2M_p^2}
		f_{1T}^q(x,k_T^2).
\end{equation}
We find for the symmetric part of the cross section for $W^-$ production
\begin{equation}\label{sigmaSW}
\frac{d\sigma^S}{dl_T dY_l d\phi_l} =
\frac{g^4 |V_{CKM}^{ud}|^2 l_T^3}{48 (2\pi)^2 s}
\int dY_{\bar l}
\frac{F}{D},
\end{equation}
and for the antisymmetric part
\begin{equation}\label{sigmaAW}
\frac{d\sigma^A}{dl_TdY_ld\phi_l} = 
\frac{g^4 M_p^2 |\mathbf{S}_T|^2 |V_{CKM}^{ud}|^2 l_T}{96 (2\pi)^2 s}
\int dY_{\bar l}
\bigg\{ \frac{A}{D^3} F^0 + \frac{B}{D^3} \big[ F^C \cos 2\phi_S^l 
 - F^S \sin 2\phi_S^l \big] \bigg\},
\end{equation}
where 
\begin{equation}
\begin{aligned}
A 	=& 150 l_T^8-32 l_T^6
	M_W^2-12 l_T^4 M_W^4+M_W^8-28 l_T^4 M_W^2
	\Gamma_W^2+2 M_W^6 \Gamma_W^2+M_W^4 \Gamma_W^4
\\
	&+4l_T^2 \Big[58 l_T^6-9 l_T^4 M_W^2+M_W^4 
	\left(M_W^2 + \Gamma_W^2\right)-2
	l_T^2 \left(3 M_W^4+5 M_W^2 \Gamma_W^2\right)\Big]
	\cosh[Y_l-Y_{\bar l}]
\\
	&+4 l_T^4 \Big[26 l_T^4-3 M_W^2
	\big(M_W^2+\Gamma_W^2\big)\Big] \cosh[2 (Y_l-Y_{\bar l})]
	+ ( 24l_T^8 + 4 l_T^6 M_W^2 )
	\cosh[3(Y_l-Y_{\bar l})]+2 l_T^8 \cosh[4 (Y_l-Y_{\bar l})],\\
B	=& 130 l_T^8+32 l_T^6 M_W^2+16 l_T^2 M_W^4
	\left(M_W^2+\Gamma_W^2\right)-M_W^4
	\left(M_W^2+\Gamma_W^2\right)^2 
	-4 l_T^4 \left(15 M_W^4+11 M_W^2 \Gamma_W^2\right)
\\
	&+4 l_T^2 \Big[54 l_T^6+9 l_T^4 M_W^2 +
	3 M_W^4 \left(M_W^2+\Gamma_W^2\right)-2 l_T^2
	\left(9 M_W^4+7 M_W^2 \Gamma_W^2\right)\Big]
	\cosh[Y_l-Y_{\bar l}]
\\
	& +12 l_T^4 \Big[10
	l_T^4-M_W^2 \left(M_W^2+\Gamma_W^2\right)\Big]
	\cosh[2(Y_l- Y_{\bar l})]
	+ \left(40 l_T^8-4 l_T^6 M_W^2\right) 
	\cosh[3(Y_l- Y_{\bar l})]
	+6 l_T^8 \cosh[4(Y_l-Y_{\bar l})],
\\
D	=& 6 l_T^4 - 4l_T^2 M_W^2 + M_W^4 + M_W^2 \Gamma_W^2 
	+ \left(8 l_T^4-4 l_T^2 M_W^2\right) \cosh[Y_l-Y_{\bar l}]
	+ 2 l_T^4 \cosh[2 (Y_l-Y_{\bar l})]
\end{aligned}
\end{equation}
and 
\begin{equation}\label{Fdef}
\begin{aligned}
F	&= e^{Y_l-Y_{\bar{l}}} f_1^d(x_1) f_1^{\bar{u}}(x_2)+ 
       e^{Y_{\bar{l}}-Y_l} f_1^d(x_2) f_1^{\bar{u}}(x_1),\\
F^0	&= e^{Y_l - Y_{\bar l}}
	\Big[f_{1T}^{\perp d(1)}(x_1)f_{1T}^{\perp \bar{u}(1)}(x_2)
	- g_{1T}^{d(1)}(x_1)g_{1T}^{\bar{u}(1)}(x_2)\Big]+
	e^{Y_{\bar l} - Y_l}
	\Big[f_{1T}^{\perp d(1)}(x_2)f_{1T}^{\perp \bar{u}(1)}(x_1)
	- g_{1T}^{d(1)}(x_2)g_{1T}^{\bar{u}(1)}(x_1)\Big],\\
F^C	&= e^{Y_l - Y_{\bar l}}
	\Big[f_{1T}^{\perp d(1)}(x_1)f_{1T}^{\perp \bar{u}(1)}(x_2)
	+ g_{1T}^{d(1)}(x_1)g_{1T}^{\bar{u}(1)}(x_2)\Big]+ 
	e^{Y_{\bar l} - Y_l}
	\Big[f_{1T}^{\perp d(1)}(x_2)f_{1T}^{\perp \bar{u}(1)}(x_1)
	+ g_{1T}^{d(1)}(x_2)g_{1T}^{\bar{u}(1)}(x_1)\Big],\\
F^S	&= e^{Y_l - Y_{\bar l}}
	\Big[f_{1T}^{\perp \bar{u} (1)}(x_2)g_{1T}^{d (1)}(x_1)
	- f_{1T}^{\perp d(1)}(x_1)g_{1T}^{\bar{u}(1)}(x_2)\Big]+
	e^{Y_{\bar l} - Y_l}
	\Big[f_{1T}^{\perp d(1)}(x_2)g_{1T}^{\bar{u}(1)}(x_1)
	- f_{1T}^{\perp \bar{u}(1)}(x_1)g_{1T}^{d(1)}(x_2)\Big].
\end{aligned}
\end{equation}
With the use of the expressions for the cross section in Eqs.\ \eqref{sigmaSW} and \eqref{sigmaAW}, 
the spin asymmetries, as defined in Eq.\ \eqref{ATTdef}, can be written as
\begin{equation}
\begin{aligned}
A_{TT}^0(l_T)	&= \frac{|\mathbf{S}_T|^2 M_p^2}{2l_T^2} 
		\frac{\int dY_{\bar l} \frac{A}{D^3} F^0 }
		{\int dY_{\bar l}\frac{F}{D}},\\
A_{TT}^{C,S}(l_T)	&= \frac{|\mathbf{S}_T|^2 M_p^2}{\pi l_T^2} 
		\frac{\int dY_{\bar l} \frac{B}{D^3} F^{C,S}}
		{\int dY_{\bar l} \frac{F}{D}}.\\
\end{aligned}
\end{equation}
The results are easily modified for $W^+$ production by substituting $\bar{u}\to\bar{d},d\to u$ in Eq.\ \eqref{Fdef}, substituting $l_T \to \bar{l}_T, \phi_l\to \phi_{\bar l}$ in all expressions and integrating over $Y_l$ instead of $Y_{\bar l}$ in the cross sections and asymmetries.
As a cross-check of the approximation method employed here, we calculated the single spin asymmetry $A_N$ in $W$ production and found reasonable agreement with the results in Refs.\ \cite{Kang:2009bp,Metz:2010xs} taking into account that different functions were used.

\begin{figure}[htb]
\subfigure[\ $Y_l=0$]
{\includegraphics[height=3.3cm]{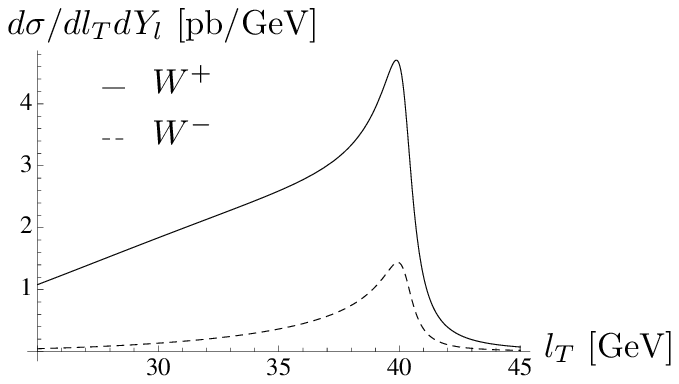}}
\subfigure[\ $l_T=40$ GeV]
{\includegraphics[height=3.3cm]{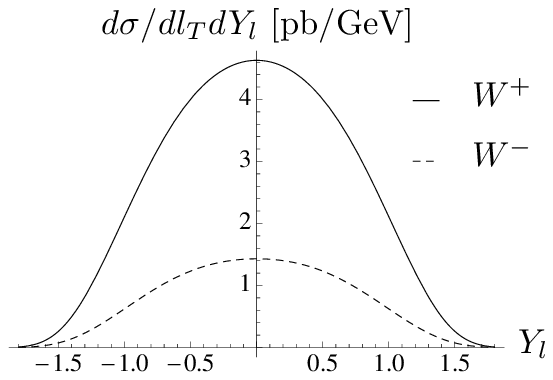}}
\caption{Differential cross section for $W$ boson production at RHIC energy $\sqrt{s}=500$ GeV.}\label{SigmaW}
\end{figure}

\begin{figure}[htb]
\subfigure[\ $Y_l=0$]
{\includegraphics[height=5cm]{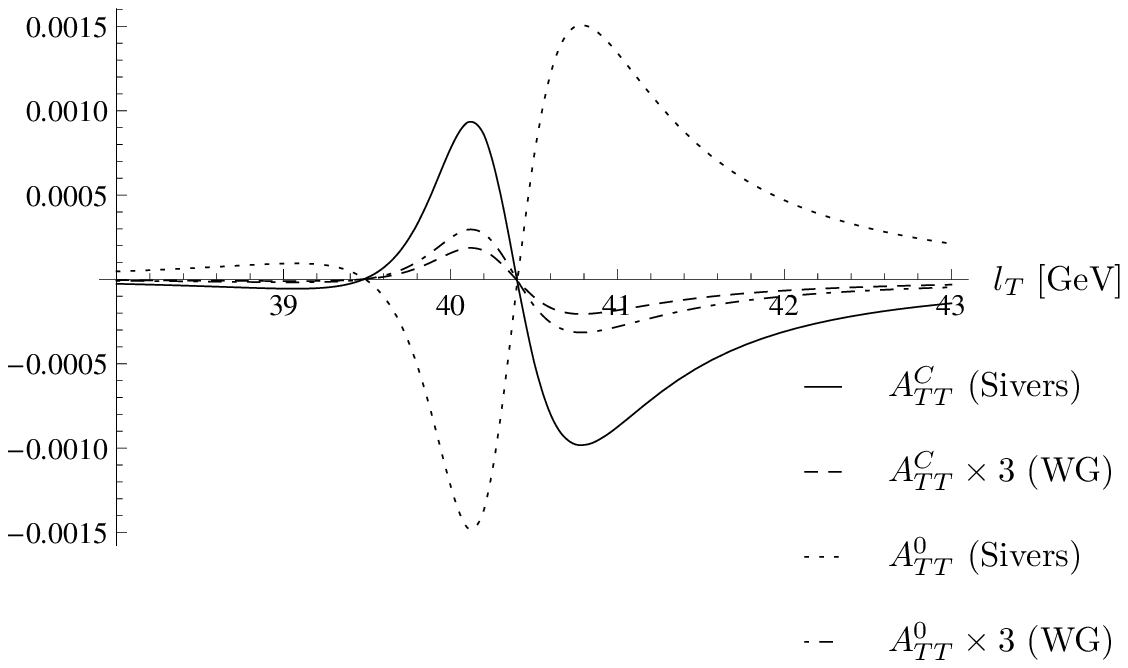}}
\hspace{1cm}
\subfigure[\ $l_T=40$ GeV]
{\includegraphics[height=5cm]{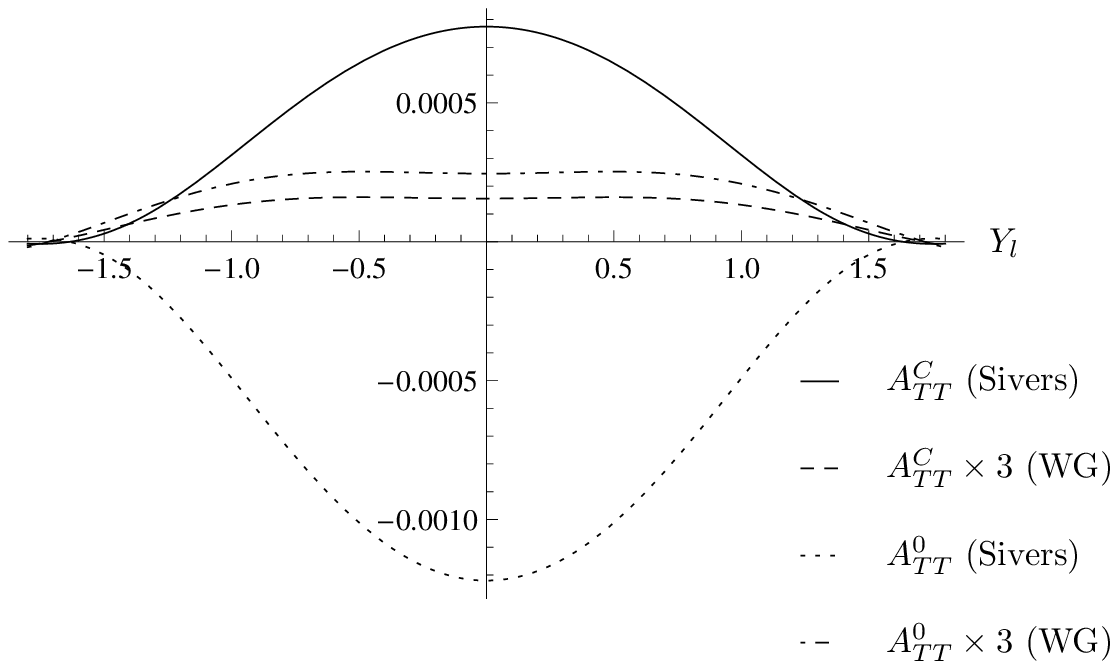}}
\caption{Contributions to $A_{TT}(l_T)$ in $W^+$ boson production from the double Sivers and Worm Gear effects at RHIC energy $\sqrt{s}=500$ GeV.}\label{ATTWp}
\end{figure}

The cross sections for $W^\pm$ production are plotted in Fig.\ \ref{SigmaW}. 
We only show the asymmetries in $W^+$ production in Fig.\ \ref{ATTWp}, because they are largest. 
The maximal asymmetry is near resonance and reaches up to $0.15\%$, which is already below the detection limit at RHIC. 
However, for a bound on a possible $W-W^\prime$ mixing it is not the differential asymmetry that is relevant, but the asymmetry in the integrated cross section. In those asymmetries the contribution at $l_T<M_W/2$ largely cancels the contribution at $l_T>M_W/2$, resulting in very small asymmetries. We find the integrated asymmetry in $W^-$ production around $10^{-7}$ and in $W^+$ production around $10^{-6}$, far below detection limits at RHIC. This confirms the expectation expressed in Ref.\ \cite{Boer:2010mc} that the background
from TMDs, is indeed negligible. 

\FloatBarrier

\section{Summary and conclusions}

We calculated the transverse momentum dependent
double transverse spin asymmetries in the laboratory frame for Drell-Yan and $W$ production arising from the
Sivers effect and from the Worm Gear distribution function $g_{1T}$ within transverse momentum dependent factorization.
Those asymmetries were previously calculated only in the Collins-Soper frame where they are independent of the lepton azimuthal angle.
The advantage being that one can, in that frame, easily distinguish them from the asymmetry coming from transversity. In the lab frame a residual dependence on the lepton azimuthal angle of the TMD asymmetries survives and enters the double transverse spin asymmetry in exactly the same way as 
transversity does. This is in contrast to a collinear factorization approach where the effects from TMDs are absent to begin with. Therefore, a nonzero $\cos 2 \phi_{S}^l$ asymmetry $A_{TT}(q_T)$ in Drell-Yan {\em in the lab frame} is {\it a priori} not a sufficient indication of a nonzero transversity distribution. However, from what is known about the magnitudes of the Sivers and Worm Gear functions, our conclusion is that the TMD background is below the permille level. Therefore, a percent level asymmetry 
{\it can} be viewed as coming from transversity. That is an important conclusion for the RHIC spin program. 
Transversity distributions can thus safely be determined from the transverse momentum dependent double spin asymmetry in the lab frame, like for the $q_T$-integrated asymmetry, assuming of course the antiquark transversity distributions are sufficiently large. As a cross-check of the smallness of the TMD background, one can verify that the angular independent $A_{TT}^0(q_T)$ 
asymmetry that arises only from the mentioned TMD effects, is indeed much smaller.

In the leptonic decay of a $W$ boson the neutrino will go unobserved, which renders it impossible to determine the Collins-Soper frame.
In that frame a correlation between the lepton azimuthal angle and the proton transverse spin direction can solely be caused by a non-zero righthanded coupling of the $W$ boson in combination with a non-zero transversity distribution, which makes it a very suitable process for the determination of a possible $W-W^\prime$ mixing as discussed in \cite{Boer:2010mc}.
In the lab frame, however, there might again be a residual asymmetry coming from the double Sivers or WG effects. 
They can lead to a nonzero result in $W$ production,
which could be mistaken for physics beyond the Standard Model or simply spoil the opportunity to bound a possible $W-W^\prime$ mixing. We obtained numerical estimates for the sizes
of the asymmetries at RHIC and found that they are far below the detection limits.
This means that the background from the double Sivers and Worm Gear effects
is negligible and does not hamper
the investigation of the complex mixing of $W$-$W'$ bosons as discussed in \cite{Boer:2010mc}.

\acknowledgments
We thank Zhongbo Kang, Piet Mulders, Alexei Prokudin, Jianwei Qiu, and Werner Vogelsang for fruitful discussions.  This work is part of the
research program of the ``Stichting voor Fundamenteel Onderzoek der
Materie (FOM)'', which is financially supported by the ``Nederlandse
Organisatie voor Wetenschappelijk Onderzoek (NWO)''. Work of A.K.  was partially supported by Regione Piemonte (Italy).

\end{document}